# DYNAMIC IDP SIGNATURE PROCESSING BY FAST ELIMINATION USING DFA

Mohammed Misbahuddin, Sachin Narayanan and Bishwa Ranjan Ghosh

Computer Networks & Internet Engineering Division, Centre for Development of Advanced Computing,

#68, E-City, Bangalore, India

**misbah,sachin,ghosh@{ncb.ernet.in}**

## ABSTRACT

*Intrusion Detection & Prevention Systems generally aims at detecting / preventing attacks against Information systems and networks. The basic task of IDPS is to monitor network & system traffic for any malicious packets/patterns and hence to prevent any unwarranted incidents which leads the systems to insecure state. The monitoring is done by checking each packet for its validity against the signatures formulated for identified vulnerabilities. Since, signatures are the heart & soul of an Intrusion Detection and Prevention System (IDPS), we, in this paper, discuss two methodologies we adapted in our research effort to improve the current Intrusion Detection and Prevention (IDP) systems. The first methodology RUDRAA is for formulating, verifying & validating the potential signatures to be used with IDPS. The second methodology DSP-FED is aimed at processing the signatures in less time with our proposed fast elimination method using DFA. The research objectives of this project are 1) To formulate & process potential IPS signatures to be used with Intrusion prevention system. 2) To propose a DFA based approach for signature processing which, upon a pattern match, could process the signatures faster else could eliminate it efficiently if not matched*

## 1. Introduction

### 1.1 Domain Background

An Intrusion Detection & Prevention System (IDPS) is a software and/or hardware designed to detect & prevent unwanted attempts at accessing, manipulating, and/or disabling of computer / information systems, mainly through a network, such as Internet. These attempts may take the form of attacks by crackers, malware and/or disgruntled employees.

An IDPS is used to detect / prevent several types of malicious behaviors that can compromise the security of a system. This includes network attacks against vulnerable services, data driven attacks on applications, host based attacks such as privilege escalation, unauthorized logins and access to sensitive files, and malware (viruses, Trojan horses, and worms). An IDPS can be composed basically of three components: 1) *Sensors* which generate security events, 2) A *Console* to monitor events and alerts and control the sensors, and 3) A *Central Engine* that records events logged by the sensors in a database and use a system of rules to generate alerts from security events received. There are several ways to categorize an IDPS depending on the type and location of the sensors and the methodology used by the engine to generate alerts. In many simple IDPS implementations all three components are combined in a single device or appliance. But an IDPS [6,7] cannot directly detect / prevent malicious activities in encrypted channels.

All the Intrusion Detection & prevention Systems can be classified as shown in fig 'a'

*Statistical anomaly based IDS-* A statistical anomaly based IDS establishes a performance baseline based on normal network traffic evaluations. It will then sample current network traffic activity to this baseline in order to detect whether or not flow is within baseline parameters. If the sampled traffic is outside baseline parameters an alarm will be triggered. Protocol anomaly detection can be used to observe various protocol fields in different protocols such as TCP, IP, Mac etc. for various values which are far off from their respective baseline values as per their respective RFC's.

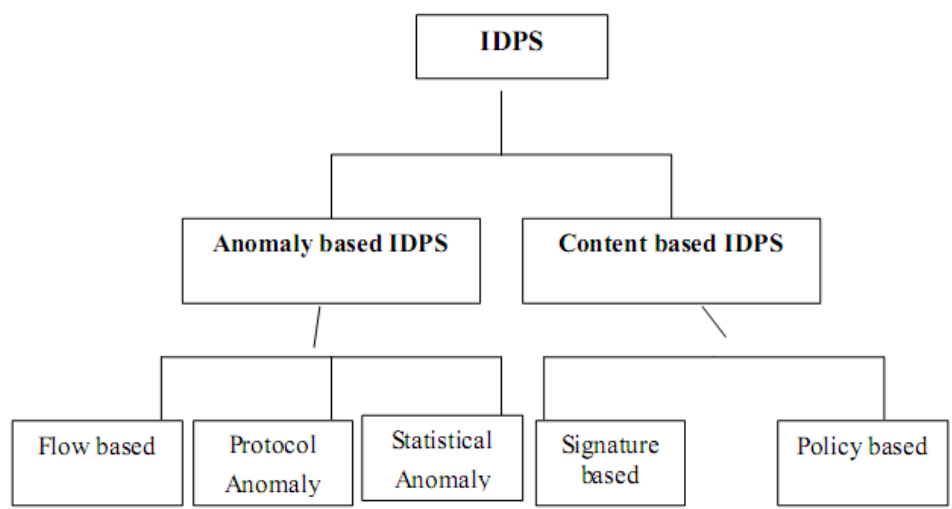

Fig 'a' - Classification of IDPS

*Signature based IDS-* Network traffic is examined for preconfigured and predetermined attack patterns known as signatures. Signatures have attributes like false-positives, false-negatives, completeness, breadth, precision, collision and recall [10]. Many attacks today have distinct signatures for vulnerabilities. In good security practice, a collection of these signatures must be constantly updated to mitigate emerging threats.

Attack analysis is the process of conducting systematic analysis on multi-stage attacks and facilitating large scale detection and visualization of security events by embedding modeling and analytical components within a more expansive security framework and basically deals with Penetration testing, Ethical hacking, Incident handling and Exploit writing.

## 1.2 The Need

Current weaknesses in IDPS include the fact that new attacks are not detected until someone has generated a rule or signature for that specific attack. Also, most attacks need a slight alteration in existing malicious code in order to bypass existing rules / signatures. Hence, new signatures are generally created manually by addressing the application vulnerabilities once the vulnerability is exploited. Some signatures may indicate which specific attack could be performed or what vulnerability the attacker is trying to exploit, while other signatures may just indicate that unusual behavior is occurring, without specifying a particular attack. So, it is essential to craft different types of signatures to identify the potential attack & try all possible combinations to negate the various mutations of that attack. To validate the IDP signatures, it's necessary to carry out the validation and attack analysis.

## 1.3 Purpose, Aim & Objective of this Research work

We have already built an IDS named as N@G (Network at Guard) [11] and we are currently developing a hardware based IPS named GYN (Guard Your Network) [12] for which we need to have our own proprietary signatures. Since the current signatures which we have been using in our IDS are from public domain (SNORT signature set[9]), we need to verify and validate these signatures for using it with our IPS. We also need to provide an improved and customized IPS for which we need to perform attacks against various exploits in order to identify its impact on the network and the victim system. Moreover, our objective in this project is to reduce the signature processing time for which we have proposed and implemented a methodology which reduces the processing time using Fast Elimination method using DFA [3]

In this paper, we are discussing two methodologies for signature processing
1) In the first methodology, RUDRAA, we process, verify & validate the potential signatures to be used with IPS using the standard signature processing method.
2) The second methodology DSP-FED is proposed by us in which we aimed at processing the signatures in less time with our proposed fast elimination method using DFA.

The rest of the paper is organized as follows:

Section 2 will cover the detailed process of RUDRAA, Section 3 will cover the methodology of DSP-FED, and section 4 will give a conclusion.

## 2. RUDRAA – an intRUsion Detection & pRevention signAture formulAtion process

This project is intended to formulate IDP signatures and provide a mechanism for code improvement and packaging for IDPS. Since signatures are the soul of an IDPS, it becomes essential to accurately formulate the signatures. A lot of signatures are available in public domains which are not verified / validated for an IPS. So, in order to do so, the process of formulating the signatures can be divided into three phases:

*Signature identification* involves identification of potential IPS candidate signatures targeting applications/class based on their relevance and severity which have publicly available references like CVE[1,5], NESSUS[4], BUGTRAQ[8] and others.

*Signature verification* is the phase where the formulated IPS Signatures are evaluated for the appropriate fields related to headers, protocols, payload, state based, and action types. This involves header analysis, payload analysis, state-based analysis and action type analysis.

*Signature validation and impact analysis:* This includes the Attack analysis for all IPS signatures. To perform this crucial activity, setup of separate attack analysis lab is essential. Attack analysis lab needs to be equipped with all widely used open source attack/exploits. Since Signature formulation is a continuous activity, the above mentioned three phases would continue in future also. The outcome of the above activities is a set of formulated IDP signatures.

### 2.1 Alternatives

Almost all IDP solution providers do have the signatures formulation mechanism in place, such as, Cisco, iPolicy, real secure, etc... Even some open source projects exist which formulate signatures on regular basis. All of these agencies do not publicize their IPS signatures. SNORT IDS signatures are publicly available, but all of these IDS signatures can not be fed into our IPS system without verification and validation.

### 2.2 Approach / Methodology

To achieve the objectives of this project, we approached the problem statement categorically in the following way:
**Signature formulation:** As discussed in the previous section, this process comprises of three phases, I) Signature identification II) Signature verification and III) Impact Analysis. To do this we took the Snort's 2.8.1 [9] signature set which counted to 13300+ IDS signatures. The signature formulation process is depicted in Fig 1.

## Fig 1: Signature Formulation Process

**I) Signature Identification**- In this phase we filtered the 13300+ Snort signatures based on (a) Targetted application relevance, (b) Severity [1,5] (signatures having severity greater than 6) and (c) Time (signatures notified after year 2000). This is shown in figure 1 as the output of S1 & S2. The details process flow chart for this step is shown in below in Fig 1.1

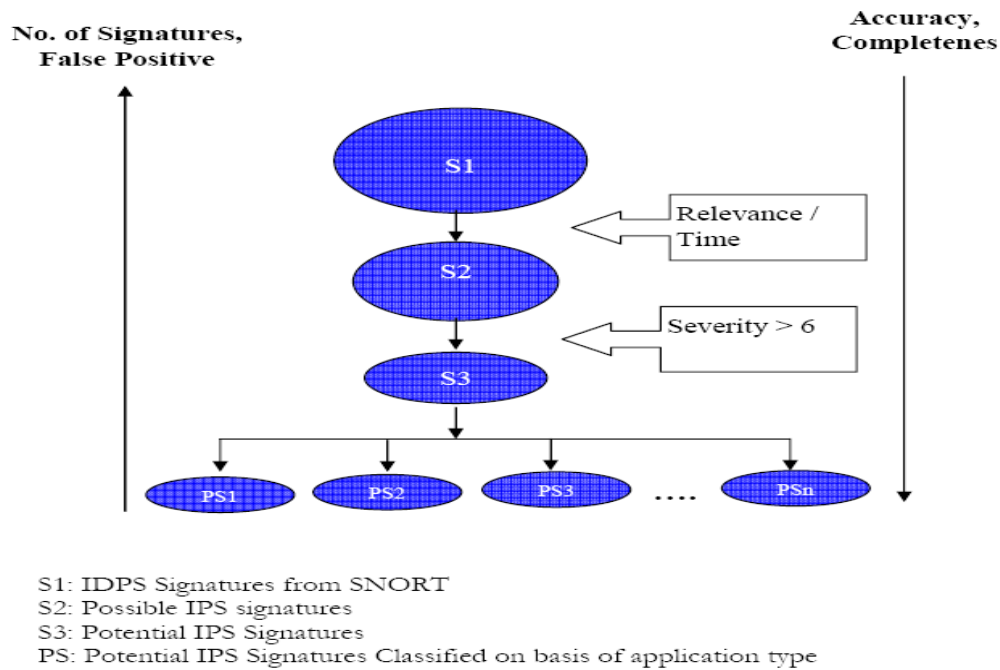

S1: IDPS Signatures from SNORT
S2: Possible IPS signatures
S3: Potential IPS Signatures
PS: Potential IPS Signatures Classified on basis of application type

**Detailed Process Explained**

1) We have taken SNORT's signature set in which we found thirty .rules files. Each file consists of certain number of signatures. During processing we take as input one .rule file at a time.
2) Each signature in each rule file is then picked up and checked if signature is valid with the timeline. If the signature is found to be invalid it is sent to a file called Disable.out set. Each such invalid signature is kept in this file.
3) If the signature is valid as per the timeline then all the CVE's associated with this signature is identified.
    a. For each CVE , the CVE ID is checked. If a CV ID is found then the corresponding severity is fetched from the database. If the associated CVE is not found then it is kept in candidate.out signature set.
    b. Once the severity is fetched from DB it is now checked whether any new CVE is published, if yes then the corresponding CVE link from NVD site is fetched and saved in the local directory
        i. Now run the parser to retrieve the severity associated with CVE-ID
        ii. Update the database with the above retrieved information
    c. Now check the severity level, if the severity is >=6 , put the signature in signature.out signature set, else check for other CVE – Ids
If all the CVEs of a signature is checked then go to the next signature

**Our Results**

We first filtered the potential IPS candidate signatures targeting applications/classes based on Application/Class Relevance, Time (after year 2000, process S1) and Severity ranging from high (80%) and medium (20%) (Process S2). The applications considered for filtering are HTTP, DNS, SMTP, SQL, RPC, POP3, Telnet, R-Services, NetBIOS, Finger, Chat, Backdoor, Attack Response and Exploit based vulnerabilities (These potential candidate signatures will have references in CVE or NESSUS or BUGTRAQ, etc. and all high severe signatures will not have the CVE reference). The resultant set after filtering according to S1 & S2 (relevance & time) contained 5480 out of 13,350 IDP signatures. This set of 5480 signatures is now subjected to filtering based on action type written in the signature. If the action is Drop packet then these signatures are considered potential IPS signatures. This step yielded 3201 candidate signatures. So, at the end of identification phase we had 3201 candidate IPS signatures.

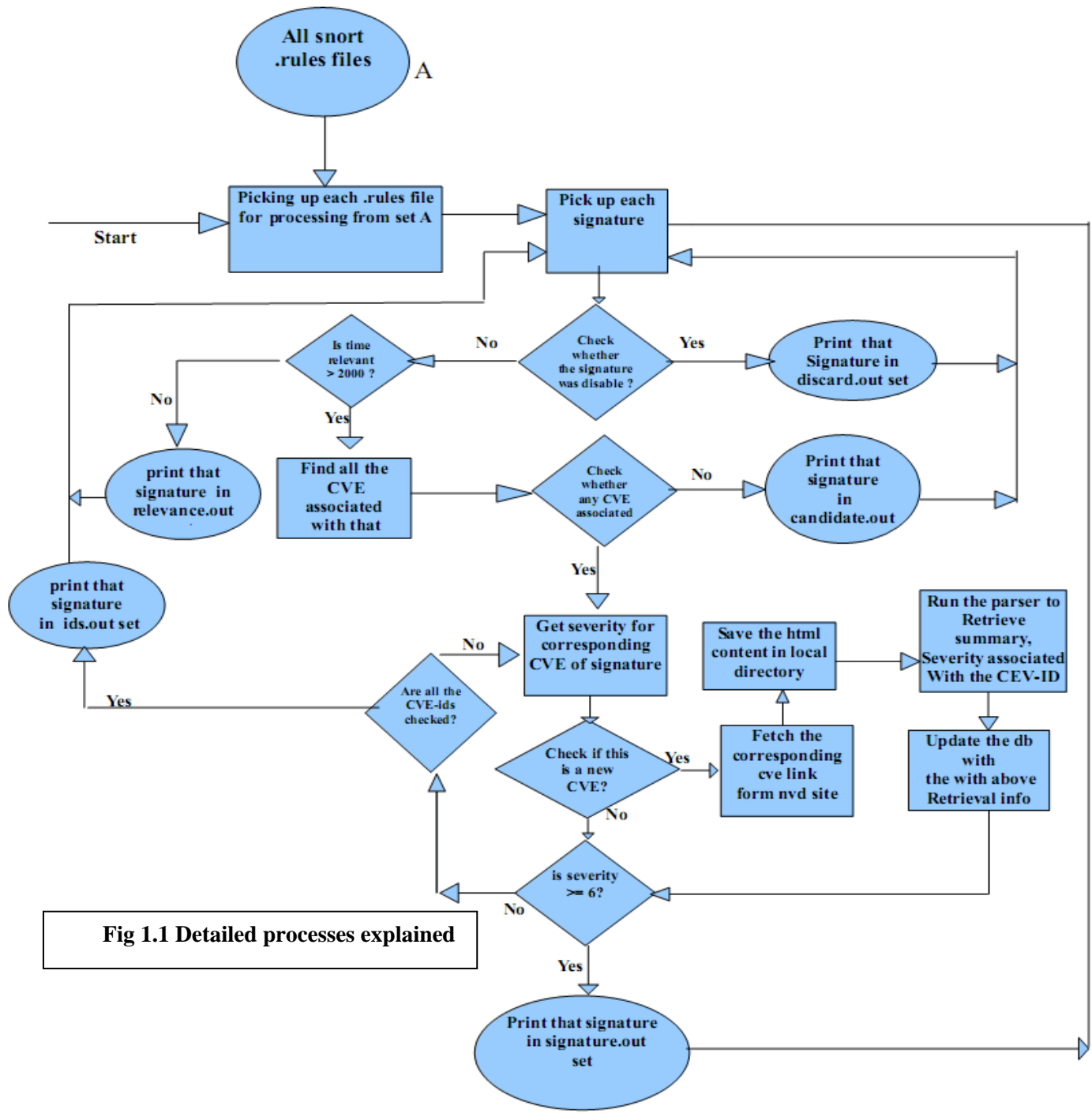

Fig 1.1 Detailed processes explained

**11) Signature Verification-**

The signature verification and validation is done by doing the following four analyses:
**Header analysis**:

This includes verification of the protocol field's value under which a signature falls, such as ARP, ICMP, IP, TCP, UDP etc...

Information about the IP (external/home network). This field has to be properly verified because the IP field value reveals the information about direction of packet transmission. IP field can have different values like, EXTERNAL_NETWORK, HOME_NETWORK, IP_RANGE and any.

Port field in signature specifies about the destined application for which the signature has formulated. This field may be assigned with different value like HTTP_PORT, SMTP_PORT, 80, 21, range (40001:56000) and any.

Other protocol related fields like flags, Type of service, IP Identification number, various sequence and acknowledgment number, TCP window size, icode, itype, icmpid etc. has to be verified for signature.

**Payload analysis**: Most of the signatures have content (payload) or value like 'uricontent' which can be either in the plain text or in regular expression form. To carry out Deep packet and content analysis, the content and uricontent fields are supported by other fields such as offset, depth, byte jump, meta data, case or nocase, distance, within, etc

**Sate based analysis**: This includes the verification of fields like, Flow information, connection state, and session information.

**Action type analysis**: After successful matching of each signature, a specific action has to be associated with it. Like DROP, LOG, ACCEPT. So, based on the severity and impact of attack pattern (signature) the action type value has to be properly assigned and verified further. Based on the action type analysis we categorized the 3201 signatures as 1294 'IPS DROP' and 1907 'IDS ALERT' signatures respectively.

**III) Impact Analysis**- This phase includes the Attack analysis for all candidate IPS signatures which is the output of signature verification phase. To perform this crucial activity, we have setup a separate attack analysis lab. The architecture of this attack lab is shown in fig 2.

We have categorized the systems as 12 Victim systems, one IDP sensor running on high end server with dual interface card. We also used 'n' (=20) attacker systems for performing various attacks based on the type of vulnerability. All these systems are connected with two switches respectively. All the systems in the attack lab network were configured with various operating systems such as Windows XP, 2000, Fedora 9, Red Hat Linux, Debian etc. These systems are installed with all widely used open source attack/exploits softwares. The different tools that were used to perform the Impact analysis are: Nemesis, Nmap, Nessus, Metasploit Framework, Wireshark, Tcpdump, Netcat, Nikto, Paros proxy, Whisker, IDSwakeup, etc…
The activities involved in Impact analysis includes:

1) Identification of Vulnerability in the system as specified by the formulated IPS signatures.
2) Exploiting the vulnerability by using scripts, codes or exploits available in the open domain

**Fig.2 Architecture of Attack Lab**

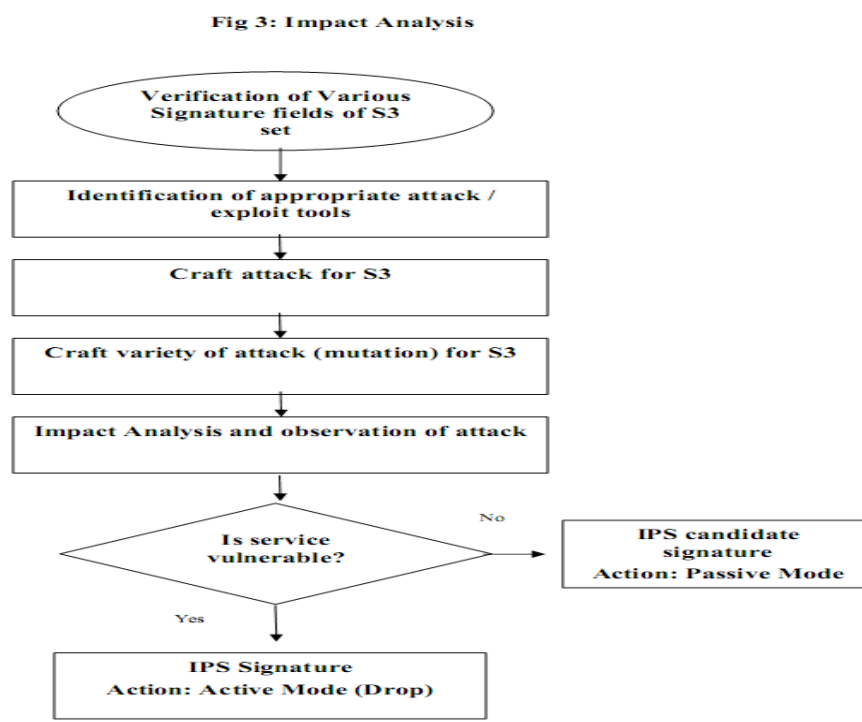

Fig 3: Impact Analysis

3) Observation and analysis of the impact of the crafted attack using various walk scenarios.
4) Decision on the action relevant for the caused impact.

In addition, the packet captures and the impact analysis of each attack is recorded for input to the IPS project. The detailed Impact Analysis activities are shown in Fig 3.

## 3. Dynamic Signature formulation by Fast Elimination using DFA

In this section, we propose a Fast Elimination approach using DFA.

*Working Methodology*

In this method each individual Snort signature is taken and its protocol is evaluated. The protocol field of the SNORT signature acts as the starting point (initial state) for our approach. Snort signatures clearly state which transport layer protocol (tcp, udp or icmp) packet they target so we can easily classify all SNORT signatures based on the transport layer protocol. This acts as the first state of DFA. Now we will discuss all the states below. During the processing the next field that comes is the source IP address and source port no. So, the next state in the DFA is the source IP. This source IP can be a specific IP or could be for a particular network which can be considered as external. The third state of the DFA is the source port number, the source port number acts as the identification for the application being targeted by the signature. Fourth state is the destination IP address which can be specific (specific target) or web servers (http servers) of the network. The fifth state is the destination port which specifies the application being target. This can be http ports or any other ports as targeted by the exploit and used by the snort signatures to identify the signatures. The sixth state is the various contents specific to the signature. It should be clear that a signature may have multiple content fields so depending on the number of the content fields the states in the DFA may increase. The penultimate state is the reference. This reference, points to various databases like Bugtarq or CVE. The final state is the SNORT ID which identifies each signature.

When the DFA is populated / generated then each signature gets appended to this DFA with the protocol field being the first state of the DFA. There would be different DFA for tcp, udp and icmp. When a packet comes for a match, the packet is matched to this signature DFA and the deeper it matches more accurate it becomes.

During the packet matching, the deeper the packet matches to the signature DFA, the signatures attached to higher nodes of the dfa are eliminated from the match leaving a lesser number of signatures and reducing matching time. The algorithm works on the principle of DFS. The advantage of this approach is that if the arriving packet belongs to tcp then none of the signatures of udp or icmp are involved in the matching process and are eliminated. Similarly traversing deeper into the DFA the signatures attached to higher nodes to that node are also eliminated leaving fewer signatures to match. Thus the elimination time & processing is reduced.

*Algorithm to Generate / Populate the DFA*

Check protocol if prot X
    if same src ip then go to nxt step else attach src ip
    if same src port then go to nxt step else attach src port
    if same dest ip then dest ip go to nxt step else attach dest ip
    if same dest port then go to nxt step else attach dest port
    if same then go to nxt step else attach content1/uricontent1
    if same then go to nxt step else attach node content2/uricontent2
    ------
    ------

   if same then go to nxt step else attach node contentN / uricontentN
   attach nxt node ref
   attach nxt node sid.
**Exit**

*Algorithm to Check / Match packet to sig*

Check protocol if prot X
   check if same src ip then go to nxt node; eliminate sigs attached to previous higher node
   check if same src port then go to nxt node; eliminate sigs attached to previous higher node
   check if same dest ip then go to nxt node; eliminate sigs attached to previous higher node
   check if same dest port then go to nxt node; eliminate sigs attached to previous higher node
   check if same content1/uricontent1 then go to nxt node; eliminate sigs attached to previous higher node
   check if same content1/uricontent1 then go to nxt node; eliminate sigs attached to previous higher node
   ---
   ---
   Check if same contentN / uricontentN then go to nxt node; eliminate sigs attached to previous higher node
   Fire alert
**Exit**

*Example using a Signature*

Alert tcp EN P1 → HN P2(msg "xyz"), flow: from-server, content " "within, dictionary, metadata: IPS-Policy-Drop, reference: CVE , BID, SID: )

*Fig 4: DFA for ICMP, TCP & UDP*

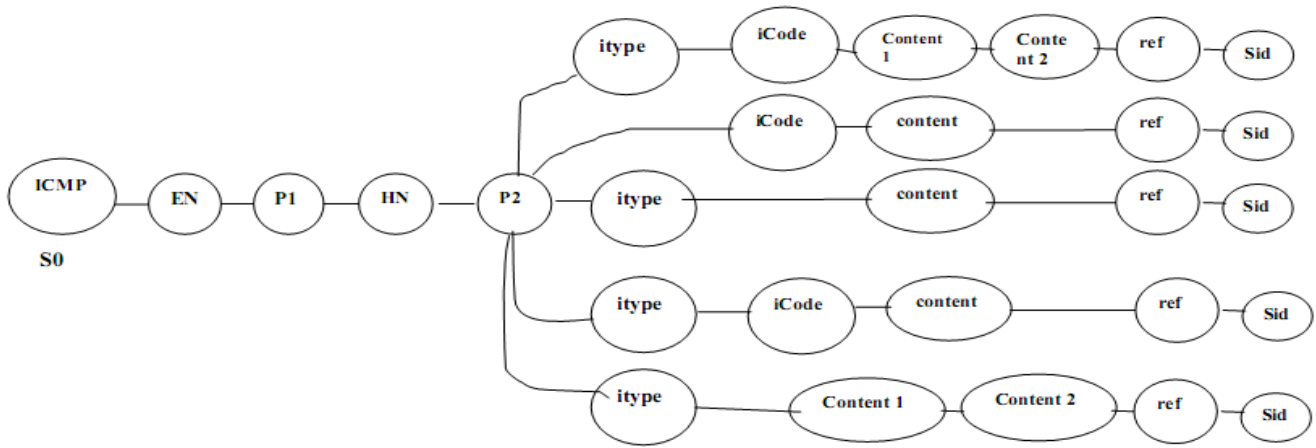

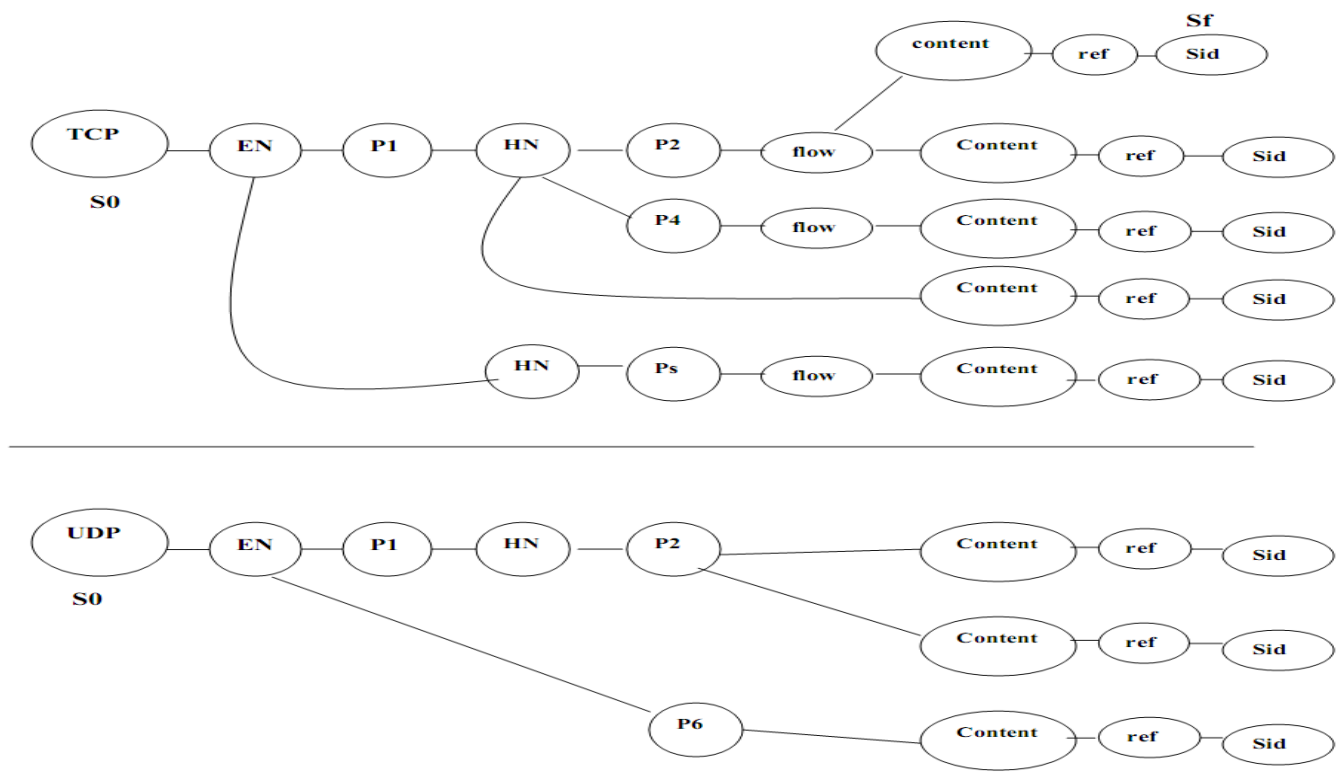

## 4. Results of DFA based approach

Using DFA based approach, we found that the time taken to eliminate a pattern against a signature is significantly reduced during the population of the DFA. Individual DFA's are generated for each protocol such as ICMP, TCP, UDP etc. which made the process of elimination easier against network traffic patterns. The novelty in this approach is derived from the elimination of signature attached to high nodes as we traverse deeper into the DFA.

To test the effectiveness of this approach, The test set was created by selecting a set of 100 signatures each for ICMP, TCP & UDP protocols. These 3 test sets $S_{icmp}$, $S_{tcp}$ & $S_{udp}$ were taken and separate DFAs were generated. The time required for each individual match was clocked by injecting packets crafted for individual signatures along with random packets. During this it was observed that our proposed approach tool less than half of the time taken by normal signature processing technique.

## 5. Conclusion

We discussed two methodologies we adapted in our research effort to improve the current Intrusion Detection and Prevention (IDP) systems. The first methodology RUDRAA is for formulating, verifying & validating the potential signatures to be used with IDPS. The second methodology DSP-FED is aimed at processing the signatures in less time with our proposed fast elimination method using DFA. Hence we found that our DSP-FED method is time efficient for signature processing.

### ACKNOWLEDGEMENTS
Our special thanks to Rajiv Ranjan who has put all his efforts to initiate this research work and giving it a shape. We also thank C-DAC for providing us with all the resources and environment to carry out this research work.

## Biography of Authors

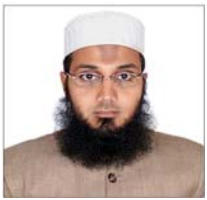

**Mohammed Misbahuddin** received his B.Tech (Engg.) in Computer Science & Engineering from K.B.N. College of Engineering, Gulbarga University, Gulbarga, Karnataka, India in 2001; M.Tech in Software Engineering from JNTU, Anantapur, India in 2006. He is presently pursuing Ph.D. in the area of Network Security from JNTU, Hyd, India. His areas of interest include Authentication & IDM, PKI and other areas of Network Security and Image Processing. He is presently working as Senior Staff Scientist in Center for Development of Advanced Computing (CDAC), Electronic City, Bangalore, India since January 2008.

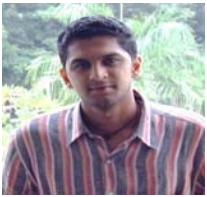

**Sachin Narayanan** is B.Tech in Computer Science & Engg. from University of Calicut. He has been working for last 2.5 years in Computer Networks and Internet Engineering Division at C-DAC (formerly NCST) #68, Electronics City Bangalore-100, India. He has been working on IDPS signature formulation for the in house products like N@G and GYN being developed by C-DAC. His areas of interest include Vulnerability Analysis and Penetration Testing.

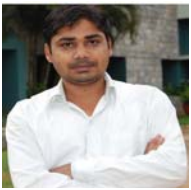

**Bishwa Ranjan Ghosh** is currently working as Staff Scientist in the Computer Network and Internet Engineering Division at C-DAC Bangalore since July 2007. He completed his B.Tech in Information Technology from College of Engineering and Technology, Bhubaneswar in 2005. Before that, he completed Intermediate of Science from TATA College, Chaibasa in Jun 1999. His research interests are in all areas of Computer Network, with special interest in network security.